\documentclass[]{elsarticle}
\usepackage{geometry}                		% See geometry.pdf to learn the layout options. There are lots.
\usepackage{graphicx}				% Use pdf, png, jpg, or eps§ with pdflatex; use eps in DVI mode
\usepackage{amssymb}
\usepackage{natbib}
\usepackage{float}

\newcommand{\degree}{^{\circ}}

\begin{document}

\begin{frontmatter}

\title{Meteorites from Phobos and Deimos at Earth?\\{\it\normalsize Accepted by Planetary and Space Science}}
\author[Lon,CPSX]{P. Wiegert} 
\author[Lon,CPSX,Vie]{M. A. Galiazzo} 
\address[Lon]{Department of Physics and Astronomy, The University of Western Ontario, London, Ontario, N6A 3K7, Canada}
\address[CPSX]{Centre for Planetary Science and Exploration (CPSX), London, Ontario, N6A 3K7, Canada}
\address[Vie]{Institute of Astrophysics, University of Vienna, Turkenschanzstr. 17, A-1180 Vienna, Austria}

%==============================================
\begin{abstract}
We examine the conditions under which material from the martian moons
Phobos and Deimos could reach our planet in the form of meteorites.
We find that the necessary ejection speeds from these moons (900 and
600 m/s for Phobos and Deimos respectively) are much smaller than from
Mars' surface (5000 m/s). These speeds are below typical impact speeds
for asteroids and comets (10-40 km/s) at Mars' orbit, and we conclude
that the delivery of meteorites from Phobos and Deimos to the Earth can
occur.

\end{abstract}

\end{frontmatter}

\section{Introduction}
Meteorites are solid interplanetary material that survives its passage
through the Earth's atmosphere and arrives at the ground. Most
meteorites originate from minor bodies, but a few arrive from planetary
bodies.  The interchange of material between the Mars and Earth is now
well-established, both from the point of view of the ejection of
material from the martian surface \citep{heameliva02}, as well as of
the orbital dynamics of Mars-Earth transfer \citep{glaburdun96}. At
this writing fifty meteorites from Mars are recognized among the
world's meteorite collections (The Meteoritical Bulletin
Database\footnote{www.lpi.usra.edu/meteor/}). Here we examine one
remaining open question in this field, and that is whether material
might arrive at the Earth from the martian satellites.

This study is partly motivated by the claim that the Kaidun meteorite
may have come from Phobos \citep{iva04}. This meteorite is largely
made up of carbonaceous chondrite material, and spectral analysis
suggests that the surface properties of Phobos and Deimos are
bracketed by outer main-belt D and T type asteroids
\citep{rivbrotri02}, the latter of which have been linked to
carbonaceous chondrites \citep{beldavhar89}. However, the Kaidun
meteorite is unusual in that it contains a wide variety of fragments
of other types \citep{iva04}.  For a meteorite to be so varied, it
would have to be formed in a region where impacts with a range of
asteroidal materials can occur, and from which escape to Earth is
dynamically possible. The martian moons certainly fit these
requirements, though they are hardly unique in this respect as most
main belt asteroids of any size do as well. We note that Phobos is
thought to have compositional variations on scales of hundreds to
thousands of meters \citep{murbrihea91, piemurtho14,baslorshi14},
consistent with a mixed composition. In any case, we do not argue here
either for or against the origin of the Kaidun meteorite from a
martian moon, but simply examine the dynamical processes by which such
a transfer might take place.

That ejecta from Phobos can escape into interplanetary space was noted
by \cite{ramhea13b} but the conditions under which this occurs was not
discussed in detail. That material released from the Voltaire impact
on Deimos might escape Mars space was discussed by \cite{naynimudr16},
though only as an aside to their study of mass transfer between the
martian satellites.  Our results differ slightly from theirs, probably
due to a different criterion for escape, and this is discussed in
Section \ref{te:nayak}.

%\cite{ramhea13b} did estimate that based on the relative
%cross-sections and escape speeds of Phobos and Mars that there are less five
%orders of magnitude more Martian than Phobian meteorites on Earth. 

%That ejecta from the Martian moons can escape more easily than from
%Mars was apparently noted in \citep{sot71} (quoted in
%\cite%p{burlamsot79} on p 24). But I haven't been able to find Soter
%1971, a PhD thesis anywhere to confirm this}

%We extend this by considering Deimos, and consider in detail the
%ejection speeds needed to escape Mars. {In particular, instead of
%  using a 60\% multiplier, we use the expected speed distributions
%  from .... We also include non-spherical ($J_2$) gravity which they did
%not, though we don't expect this to have a large effect.

The aim of this work is to determine if material ejected from either
of the martian moons by impacts might escape to reach the Earth. In
particular, we want to determine the ejection speeds required for this
to occur, and compare them to typical impact speeds at Mars to
determine whether the process is likely to occur. We do not examine
the transfer process from Mars to Earth in detail, as this has been
done by other authors \citep{glaburdun96, gla97} but simply assume
that any material that leaves near-Mars space for interplanetary space
could potentially reach the Earth. We will show that the speed
distribution of ejecta from Mars and from its moons is similar upon
leaving near-Mars space. As a result, the heliocentric orbits of
ejecta from these sources is similar, and the dynamical transfer
process from Mars to Earth should proceed in the same manner.

\section{Methods}

Particles ejected from Phobos and Deimos at different speeds are
numerically simulated under the influence of Mars, the Sun and the
other planets. One thousand particles are integrated for each ejection
speed investigated. The RADAU integrator \citep{eve85} has used with a
tolerance of $10^{-12}$. In addition, the maximum time step allowed is
0.01 Earth days (14.4 minutes or $1/30^{th}$ of Phobos' 7.65 hour
orbital period).  The $J_2$ component of the martian planetary
potential is included. The mass of Mars is taken to be of $1/3098708$
solar masses \citep{sta98}, Mars' pole to be at RA (J2000) =
317.68$\degree$, Dec (J2000) = +52.886$\degree$ \citep{jaclai14}, its
equatorial radius to be 3396.2 km\citep{arcahebow11} and $J_2 = 1.9605
\times 10^{-3}$ \citep{yod95}.

Phobos' and Deimos' orbits are taken from \cite{jaclai14} for JD
2433282.5 (1 Jan 1950 TDT, their reference date). Particles are
ejected spherically symmetrically from Phobos and Deimos at 10
positions spaced evenly around their orbits in time, to capture any
effects related to the moons' orbital phase. In practice, impacts onto
the moons are anisotropic and we expect more ejecta in the prograde
direction due to higher impact speeds on the leading side
\citep{col93}, but we are not attempting to model the amount of
material released in detail. For comparison, we also eject particles
from Mars' surface. Ten thousand particles are simulated at each
ejection speed, with the material released outwards on a set of
hemispheres located randomly on the martian surface. The effect of the
martian atmosphere is neglected.

The effect of Mars location around its heliocentric orbit is known to
be small, and impacts at aphelion are more likely since the planet
spends more time there \citep{gla97}. As a result, here we consider
only ejection of material from Mars when it is near
aphelion.\footnote{On 1 Jan 1950 TDT, Mars is near aphelion at a mean
  anomaly of $169.43\degree$.}

The martian moons are included in the simulation, their masses are
taken from \cite{jac10}, $1.08 \times 10^{-15}$ solar masses for
Phobos and $7.62 \times 10^{-16}$ for Deimos. Gravitational
perturbations by them are applied, and collisions with them are
checked for, however they are treated as spheres of their mean radius
for the purposes of collision detection. Collisions of ejecta with the
moons are not seen in any of our simulations. Certainly the
re-accumulation of ejecta by the moons would occur in practice, and may
be an important process in the resurfacing of these moons
\citep{ramhea13b}. However, given the small number of particles
simulated, the small cross-sections of the moons and the short time
scales considered, the effect of re-accumulation by the moons is
negligible for our purposes.

Collisions with Mars are recorded and are a common outcome.  The
effects of radiation pressure and Poynting-Robertson drag are ignored
here. These are important to the dynamics of small (cm-sized and
below) particles, and any collision will certainly produce vast
numbers of particles in this size range. However, for a meteorite to
survive its high-speed entry into Earth's atmosphere, it must be of
decimeter class or larger \citep{cepborelf98}, and radiation effects are
negligible for these bodies, at least on the time scales examined
here.

Simulations proceed for 10 martian years (18.8 Earth years, $\sim
20000$ Phobos periods or $\sim 5000$ Deimos periods). This was chosen
because it represents more than $10^3$ dynamical times for the ejecta,
and because the bulk of the escaping ejecta in simulations escapes on
much shorter time scales. As a test, a few simulations were run for
longer (up to 100 Earth years) but only a very few additional
particles escape over this longer time. The overall results are very
similar and are not reported on here.

As a test of the method itself, we performed additional integrations
of three of the simulations for Phobos where the results are the most
mixed, and hence the simulations most sensitive to poor
calculations. These three we the 800, 900 and 1000 m/s ejections,
described in Section~\ref{results}. These were rerun exactly the same
initial conditions but with a maximum step size reduced by half to
0.005 days and with the integration tolerance reduced to $10^{-13}$
instead of $10^{-12}$. The outcomes of the simulations were found to
be identical in terms of numbers of particles reaching each end state,
providing support for the reliability of the integration results.

The three end states of the ejecta considered here are either
collision with Mars, escape from Mars (taken to occur when it leaves
the Hill sphere, $1.08 \times 10^{6}$ km in radius), or survival on a
circumplanetary orbit. Material from Phobos and Deimos that escapes
Mars gravity well should evolve dynamically in much the same way as
ejecta from Mars itself. This process has been extensively studied
\citep{glaburdun96}, and is not duplicated here. The presence of
escaping particles in our simulation will be taken as an indication
that it is possible for material from the martian moons to reach Earth
as meteorites.

\section{Results} \label{results}

\begin{figure}
\includegraphics[width=15cm]{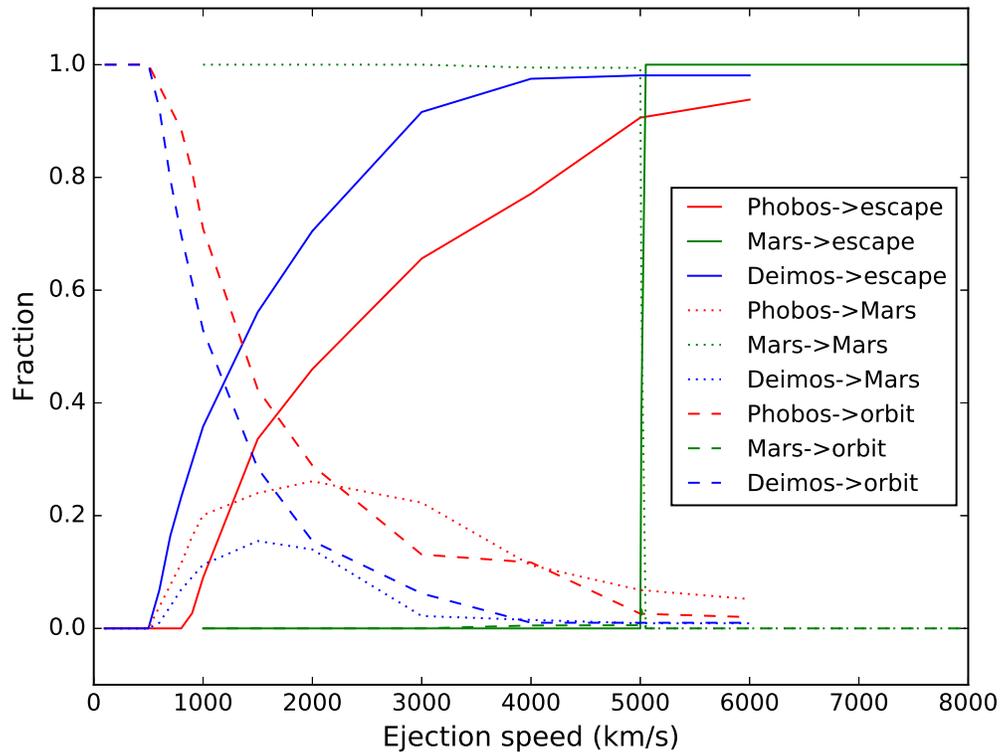}
\centering
\caption{End state fractions of particles versus ejection speed. Red
  is for Phobos, blue for Deimos and green is from Mars' surface. The
  solid lines (``escape'') indicate the number of particles which
  escape Mars' Hill sphere, dotted lines (``Mars'') are those that
  impact or return to Mars' surface, and the dashed lines (``orbit'')
  indicate those particles that are still in orbit around the planet
  at the end of the simulations. }
\label{fi:endstates}
\end{figure}

The results are summarized in Figure~\ref{fi:endstates}.  For impacts with Phobos, the lowest
ejecta speed for which escape is seen is 900 m/s. This
is much lower than the nominal escape speed from Phobos' orbital
radius from Mars, 3 km/s, because its orbital speed of 2.1 km/s
can contribute to the total for impacts on the leading surface. The process reaches 50\% escape at roughly
2500 m/s. This inner moon is harder to escape from for two reasons. First, because it is deeper in the
gravitational well of Mars; and second, because Mars' has a larger angular size,
and so impacts with the planet are more likely from simple geometry.

Deimos requires only 600 m/s for the first escapes to occur, again
from the moon's leading edge.  This ejection speed is consistent with
the difference between the local escape speed from Mars (1.9 km/s) and
Deimos' orbital speed (1.3 km/s). Escape is easier from Deimos,
beginning at speeds 30\% less than from Phobos, and ejecta reaches
50\% escape at speeds near 1500~m/s vs. 2500~m/s for the inner moon.

\label{te:nayak} The onset of escape at a 600 m/s ejection speed from Deimos differs
from the results obtained by \cite{naynimudr16} as presented in their
Figure 1, where escape can occur at speeds as low as 400 m/s. This
discrepancy seems to arise from their choice of the sphere of
influence as the boundary at which a particle is deemed to escape. The
radius of Mars' sphere of influence or sphere of activity $R_{I}$ (e.g. \cite{roy78})
\begin{equation}
  R_I = \left( \frac{M_{Mars}}{M_{\odot}} \right)^{2/5} a_{Mars}
\end{equation}
where $M$ represents the masses of Mars and the Sun, and $a$ is the
semi-major axis of Mars, is 0.579 million km and smaller than the Hill
sphere $R_{H}$ (e.g. \cite{murder99})
\begin{equation}
R_H = \left( \frac{m_{Mars}}{M_{\odot}} \right)^{1/3} a_{Mars}
\end{equation}
at 1.08 million km,  which is what we use. In our
simulations, we find that particles with ejection speeds as low as 400
m/s can reach the sphere of influence, confirming \cite{naynimudr16}'s
results in this regard, however these particles do not escape from
Mars' Hill sphere.  We verify our choice of the Hill sphere as the
appropriate boundary for escape by confirming that all of our
particles which cross this surface move out into interplanetary space
and do not immediately return to near-Mars space.  We note that the
boundary choice of \cite{naynimudr16} is unlikely to affect their
results, though they are slightly underestimating the amount of
material that remains within Mars gravity well.

From Mars' surface, the speeds needed for escape are given by the
usual escape speed from the planet, 5.0~km/s. Below these speeds,
re-impact with Mars is almost inevitable. However, at even at speeds
of 4 km/s, we note that a small fraction ($0.5\%$) of ejecta goes into
orbit around Mars, though this contribution is almost imperceptible in
Figure~\ref{fi:endstates}. Our model of ejection from Mars' surface is
not completely radial but allows ejection angles from zero to ninety
degrees from the local normal. As well, the addition of the $J_2$ term
to the potential also allows the transfer of angular momentum to the
orbits. As a result, particles ejected onto orbits near the escape
speed from Mars do not necessarily fall back to the planet's surface
but can remain in orbit.  Thus the impact events which produce martian
meteorites likely inject material into martian orbit as well, but this
has not been investigated further here.

The speeds needed for escape from the martian moons are achievable in
large impacts. Simulations of the large 9-km Stickney crater on Phobos
by \cite{aspmel93} show peak particle velocities that can reach these
speeds and higher, though only for a small fraction of the ejecta.  It
is unlikely, however, that Kaidun was ejected by the Stickney impact
itself. Stickney is expected to have formed more than 2.6 Gyr ago
\citep{schmiciva15}, though some have argued for a younger age 0.5 Gyr
\citep{ramhea15}.  In either case, the lifetime of material on orbits
crossing those of the terrestrial planets, only a few million years
\citep{glamigmor97}, is much shorter than this large crater's age. The
Kaidun meteorite was recovered as a freshly-found meteorite fall, and
is unlikely to be as old as Stickney.  Whether impacts onto from
Phobos or Deimos have released material into interplanetary space in
the last several million years is unclear, but we conclude that it is
a possibility.

\begin{figure}
\includegraphics[width=8cm]{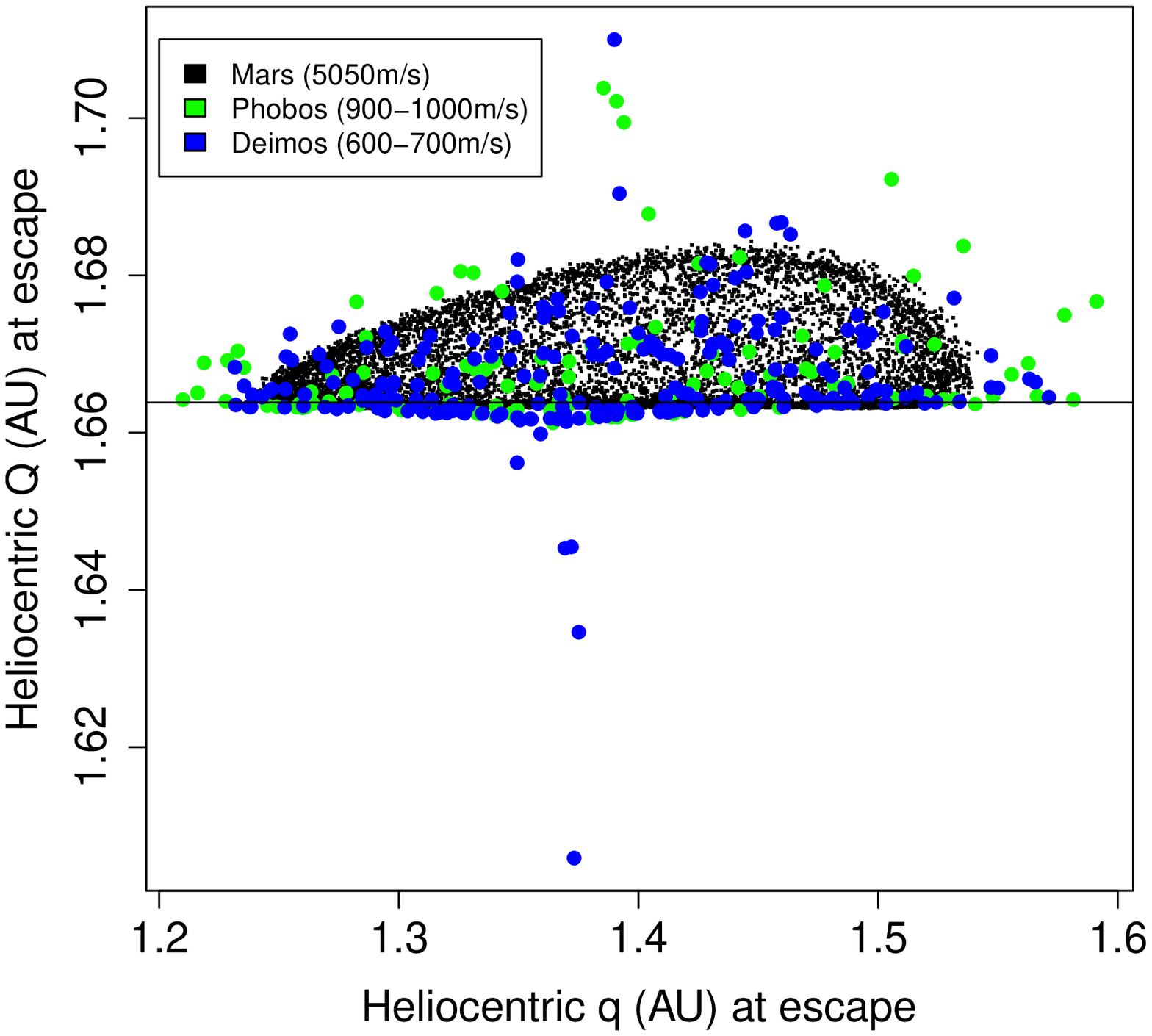}
\includegraphics[width=8cm]{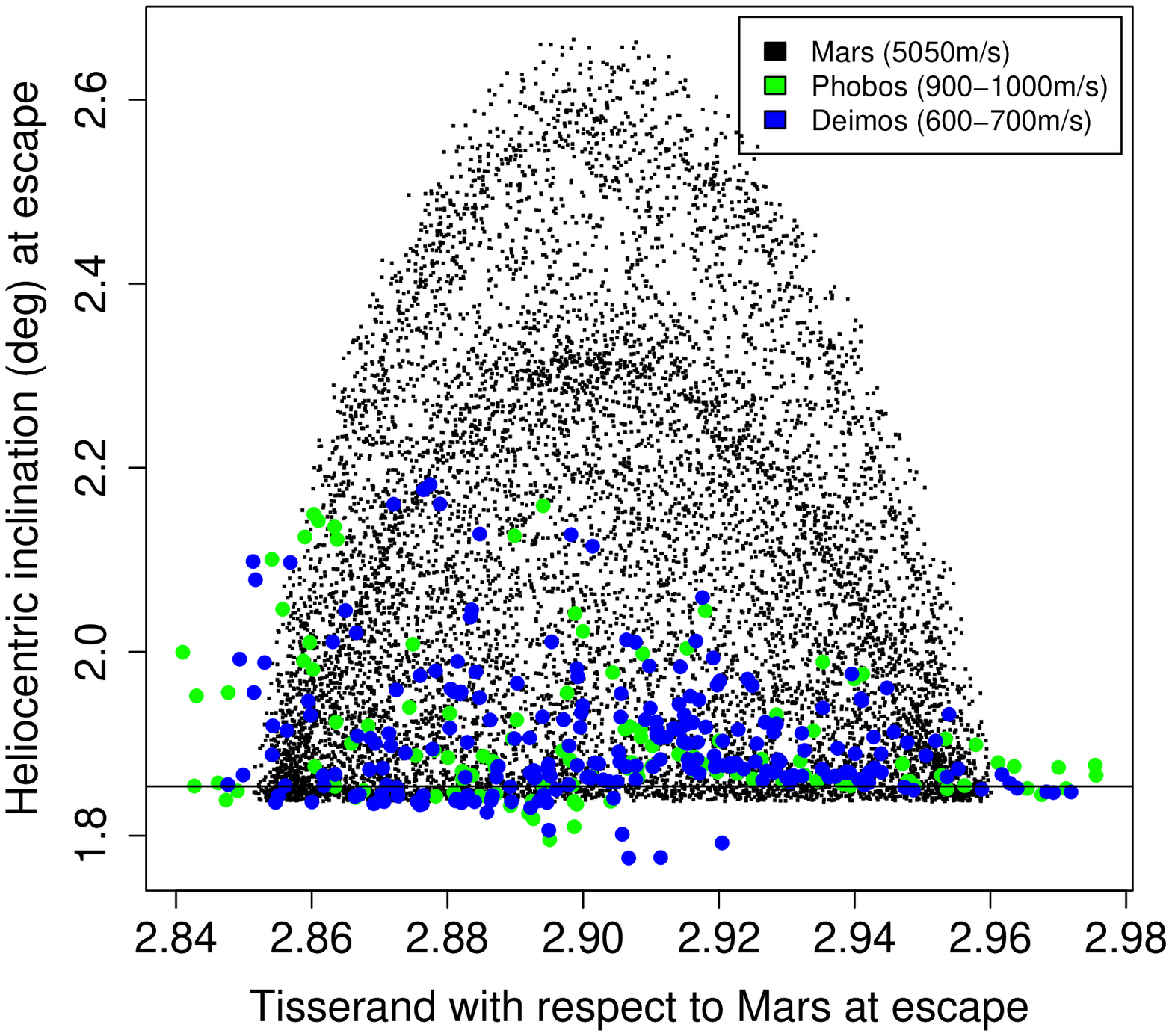}
\centering
\caption{Upper panel: Heliocentric perihelia and aphelia of material escaping from Mars' Hill sphere at the lowest launch speeds for each target. The horizontal line indicates the Mars-Sun distance at the start of the simulation. Lower panel: Heliocentric inclination and Tisserand parameter of the same material. The horizontal line indicates the inclination of Mars orbit.}
\label{fi:qQ1}
\end{figure}

The escape trajectories of material from Phobos and Deimos are similar
to that ejected from Mars. Figure~\ref{fi:qQ1} shows heliocentric
perihelion and aphelion distances for the material ejected at lowest
escape speeds from their respective targets, as well as their
inclinations and Tisserand parameters relative to Mars.. Given that
lower speeds are easier to generate than higher ones, we expect that
the bulk of the material will be released just above the minimum
threshold in most cases. The similar spreads for the different types
of material in this figures justifies the assertion that their
transfer to the Earth should occur by very similar means as well. Thus
we expect that material from Phobos and Deimos can reach Earth by the
same dynamical pathways as martian material.

An interesting feature of Fig.~\ref{fi:qQ1} is the smattering of a few
points from Phobos and Deimos at higher/lower $Q$ than is seen in the
Mars results. It arises from particles which do not immediately
escape, but which rather are injected into Mars orbit, followed by
eventual diffusion into interplanetary space through the $L_1$ and
$L_2$ points.

\section{Discussion}

The ejection of material from Phobos or Deimos into interplanetary
space, where it may be transferred to Earth, is seen at ejecta speeds
below 1 km/s: 900m/s for Phobos, 600 m/s for Deimos.

The typical asteroidal impact speed at Mars is estimated to be
$10$~km~s \citep{lefwie11, galbazdvo13}. Though these are likely the
most common impactor, comets are also possible. Long-period comets
impact at typical speeds of 40 km/s, and Jupiter-Family comets, 20
km/s \cite{chrobelup14}. These high impact speeds imply that the
ejection of material from the martian moons to escape speeds is
possible.

The escape velocity from Mars is 5.0 km/s, and the existence of
martian meteorites allows us to conclude that martian ejecta reaching
at least such speeds is produced by impacts. However, we cannot so
easily conclude that lower-speed ejecta is necessarily produced by
similar impacts on Phobos or Deimos. The speed at which ejecta is
released is strongly dependent on the target properties, and
\citep{heameliva02} for example have shown that ejecta speeds from
damaged as opposed to solid surfaces is much reduced. The nature of
the surfaces of Phobos and Deimos indicates that the conditions that
apply on Mars are unlikely to be reproduced there.

Determining the relative efficiency of meteorite delivery from Mars
and its satellites is difficult. Though arguably the speed
distribution of impacts onto Mars and its moons, as well as the
dynamical transfer from source to the Earth are similar, the
difference in the ejecta production processes from each body
complicates matters greatly.  Mars has higher gravity, an atmosphere,
a different composition, not to mention that ejection from the martian
surface was shown by \cite{heameliva02} to be enhanced by the presence
of strong target material (bedrock) on the planetary surface, all
conditions which are not applicable to the moons. Instead we ask a
simpler question: what are the implications of $\sim1$ Phobian
meteorite in our meteorite collections vis-a-vis the ejection process, and are
they reasonable? If we make the rough estimate that the bulk of the
relevant ejecta is produced by the smallest impacts capable of doing
so, the question becomes what is the smallest impactor onto
Phobos/Deimos that will produce one decimeter sized fragment?
\cite{heameliva02} found that an impact by an asteroid as 3 km in
diameter would produce $\sim 10^7$ decimeter size fragments with
escape speeds high enough for them to become martian meteorites on
Earth. Using the subscript $P$ for Phobos, a simple dimensional
estimate of the relative fraction of meteorites from Mars to those
from Phobos is
\begin{equation}
\frac{N}{N_p} \sim \left(\frac{D}{D_p}\right)^{-\alpha-1} \frac{F}{F_p}  \left(\frac{R}{R_P} \right)^2 \label{eq:ratio}
\end{equation}
where $N$ is the number of meteorites in our meteorite collections, $D$ is the
diameter of the smallest asteroid that can produce suitable fragments,
$F$ is the number of those fragments, and $R$ is the radius of the
body. The size distribution is parameterized as a power-law by
$\alpha$, which typically is used to describe the cumulative size
distribution, though we are using the differential distribution here,
hence the $-1$. The value of $\alpha$ is $\approx 1.95$ for the near-Earth
asteroid population ($e.g.$ \cite{stubin04}) and we adopt this value for the Mars
impactor population. If we solve Eqn.~\ref{eq:ratio} for $D_P$, the smallest impactor onto Phobos which
can generate one decimeter fragment at a speed high enough to escape Mars' Hill sphere, we get
\begin{eqnarray}
D_P &\sim& D \left[ \left(\frac{N}{N_p}\right)^{-1} \frac{F}{F_p}  \left(\frac{R}{R_P} \right)^2 \right]^{1/(-\alpha-1)} \\
 &\sim& 3000~\rm{m} \left[ \left(\frac{50}{1}\right)^{-1} \frac{10^7}{1}  \left(\frac{3387}{11} \right)^2 \right]^{-0.34} \\
 &\sim& 0.3~\rm{m}
\end{eqnarray}
where we have assumed a radius of 3387~km for Mars and 11~km for
Phobos. We conclude that it is reasonable that our meteorite collections would
contain 1 meteorite from Phobos if a 0.3 m impactor onto Phobos could
release one decimeter particle at a speed greater than 800 m/s.

The amount of mass released doesn't present a problem, as a high-speed
impactor might excavate a crater of whose volume exceeds its own by a
substantial factor. The volume excavated is also likely to contain
some larger pieces, and not just fines: the surfaces of the Moon, Eros
and Itokawa have volume fractions of boulders which range from less
than 1\% to 25\% \citep{micnakhir08}. But the question of whether the
large fragments could attain high enough speeds is unclear, as the
ejection of fragments of a decimeter sizes from asteroidal bodies is
not well-studied. Most impact experiments into porous/unconsolidated
targets tend to be done at centimeter impact sizes or below, and to
show low ejection speeds ($\sim 1-10$~m/s) \citep{okeahr85, ryamel98,
  houhol03, micmorhas07}. The fastest ejecta is seen to reach higher
speeds (e.g. a few hundred m/s for a 2300 m/s impact into basalt
powder \citep{har85}) but larger fragments tend to have smaller speeds
\citep{ryamel98}. There has been some simulation work done which has
shown that ejecta speeds above 1000 m/s can be achieved for
meter-sized ejecta during impacts by 8 m impactors into km-sized
porous asteroids \citep{asposthud98}; this would imply that similar
impacts on the martian moons might do the same. However, in contrast
\citep{houhol03} found lower ejection velocities, and argued that the
absence of compaction and the scale of the pore spaces in the
\citep{asposthud98} simulations might account for their higher
ejection velocities. So we can conclude that there may be
Phobos/Deimos meteorites to Earth, with the primary unknown the
distribution of ejecta speeds from impacts onto the martian moons.

\section{Conclusions}

The process of meteorite delivery from the martian moons to Earth is
examined. Two factors which increase its efficiency relative to that
of meteorites from the surface of Mars are the lowered ejection speeds
needed, and the ability of smaller impactors to launch
material. However, these are offset by the smaller cross-sections of
the moons, and the primary unknown, which is the speed distribution
for large ejecta fragments from this impacts. We conclude that it is
not unreasonable that Phobos/Deimos meteorites might exist among the
Earth's current meteorite collection, but more detailed analysis will have to await
a better understanding of the cratering process on the moons themselves.

%Consider the impact speed distribution paper in my meteor/yark paper

%Impact velocities from computer simulations \citep{ryamel98} seem to
%give speeds which are too low to allow for escape! (BUt I think these
%simulations are considered incorrect?)
%
%Velocities of ejecta from impacts into sand, particularly low angle, high vel from \citep{hersch10} show ...
%
%Crater ejecta velocity from \citep{okeahr85}
%
%Velocity distribution of large fragments \citep{vic86} implies cutoff at 1km/s but this has been superceded by ...
%
%ejecta from Mars typically hits Phobos at 2-3 km/s  \citep{ramhea13b}
%
%There is evidence for low-velocity impacts on Phobos ascribed to
%inter-moon material transfer \citep{smileeham15}

%Simulation of Stickney \citep{luccrepaj15}. But this doesn't have any information on ejection speeds

\section{References}
This work was supported in part by the Natural Sciences and Engineering
Research Council (NSERC) of Canada Discovery Grant program.  MAG was
supported by the FWF [Project J-3588-N27 “NEOs, impacts, origins and
stable orbits”].

%Formation of Phobos and Deimos by giant impact \citep{citgenida15}
%\section{More topics, to be worked into introduction, etc}
%done: Composition of surface of Phobos and Deimos \citep{piemurtho14}
%done: Ejecta from Mars makes the grooves on Phobos? \citep{ramhea13}

\section{References}

\bibliographystyle{apalike}\biboptions{authoryear}
\bibliography{Wiegert}

\begin{thebibliography}{}

\bibitem[{Archinal} et~al., 2011]{arcahebow11}
{Archinal}, B.~A., {A'Hearn}, M.~F., {Bowell}, E., {Conrad}, A., {Consolmagno},
  G.~J., {Courtin}, R., {Fukushima}, T., {Hestroffer}, D., {Hilton}, J.~L.,
  {Krasinsky}, G.~A., {Neumann}, G., {Oberst}, J., {Seidelmann}, P.~K.,
  {Stooke}, P., {Tholen}, D.~J., {Thomas}, P.~C., and {Williams}, I.~P. (2011).
\newblock {Report of the IAU Working Group on Cartographic Coordinates and
  Rotational Elements: 2009}.
\newblock {\em Celestial Mechanics and Dynamical Astronomy}, 109:101--135.

\bibitem[{Asphaug} and {Melosh}, 1993]{aspmel93}
{Asphaug}, E. and {Melosh}, H.~J. (1993).
\newblock {The Stickney impact of Phobos - A dynamical model}.
\newblock {\em Icarus}, 101:144--164.

\bibitem[{Asphaug} et~al., 1998]{asposthud98}
{Asphaug}, E., {Ostro}, S.~J., {Hudson}, R.~S., {Scheeres}, D.~J., and {Benz},
  W. (1998).
\newblock {Disruption of kilometre-sized asteroids by energetic collisions}.
\newblock {\em Nature}, 393:437--440.

\bibitem[{Basilevsky} et~al., 2014]{baslorshi14}
{Basilevsky}, A.~T., {Lorenz}, C.~A., {Shingareva}, T.~V., {Head}, J.~W.,
  {Ramsley}, K.~R., and {Zubarev}, A.~E. (2014).
\newblock {The surface geology and geomorphology of Phobos}.
\newblock {\em Plan. Space Sci.}, 102:95--118.

\bibitem[{Bell} et~al., 1989]{beldavhar89}
{Bell}, J.~F., {Davis}, D.~R., {Hartmann}, W.~K., and {Gaffey}, M.~J. (1989).
\newblock {Asteroids - The big picture}.
\newblock In {Binzel}, R.~P., {Gehrels}, T., and {Matthews}, M.~S., editors,
  {\em Asteroids II}, pages 921--945.

\bibitem[{Ceplecha} et~al., 1998]{cepborelf98}
{Ceplecha}, Z., {Borovi{\v c}ka}, J., {Elford}, W.~G., {Revelle}, D.~O.,
  {Hawkes}, R.~L., {Porub{\v c}an}, V., and {{\v S}imek}, M. (1998).
\newblock {Meteor Phenomena and Bodies}.
\newblock {\em Space Science Reviews}, 84:327--471.

\bibitem[{Christou} et~al., 2014]{chrobelup14}
{Christou}, A.~A., {Oberst}, J., {Lupovka}, V., {Dmitriev}, V., and
  {Gritsevich}, M. (2014).
\newblock {The meteoroid environment and impacts on Phobos}.
\newblock {\em Plan. Space Sci.}, 102:164--170.

\bibitem[{Colwell}, 1993]{col93}
{Colwell}, J.~E. (1993).
\newblock {A general formulation for the distribution of impacts and ejecta
  from small planetary satellites}.
\newblock {\em Icarus}, 106:536.

\bibitem[{Everhart}, 1985]{eve85}
{Everhart}, E. (1985).
\newblock {An efficient integrator that uses Gauss-Radau spacings}.
\newblock In {Carusi}, A. and {Valsecchi}, G.~B., editors, {\em Dynamics of
  Comets: Their Origin and Evolution}, pages 185--202, Dordrecht. Kluwer.

\bibitem[{Galiazzo} et~al., 2013]{galbazdvo13}
{Galiazzo}, M.~A., {Bazs{\'o}}, {\'A}., and {Dvorak}, R. (2013).
\newblock {Fugitives from the Hungaria region: Close encounters and impacts
  with terrestrial planets}.
\newblock {\em Plan. Space Sci.}, 84:5--13.

\bibitem[{Gladman}, 1997]{gla97}
{Gladman}, B. (1997).
\newblock {Destination: Earth. Martian Meteorite Delivery}.
\newblock {\em Icarus}, 130:228--246.

\bibitem[{Gladman} et~al., 1996]{glaburdun96}
{Gladman}, B.~J., {Burns}, J.~A., {Duncan}, M., {Lee}, P., and {Levison}, H.~F.
  (1996).
\newblock {The Exchange of Impact Ejecta Between Terrestrial Planets}.
\newblock {\em Science}, 271:1387--1392.

\bibitem[{Gladman} et~al., 1997]{glamigmor97}
{Gladman}, B.~J., {Migliorini}, F., {Morbidelli}, A., {Zappala}, V., {Michel},
  P., {Cellino}, A., {Froeschle}, C., {Levison}, H.~F., {Bailey}, M., and
  {Duncan}, M. (1997).
\newblock {Dynamical lifetimes of objects injected into asteroid belt
  resonances}.
\newblock {\em Science}, 277:197--201.

\bibitem[{Hartmann}, 1985]{har85}
{Hartmann}, W.~K. (1985).
\newblock {Impact experiments. I - Ejecta velocity distributions and related
  results from regolith targets}.
\newblock {\em Icarus}, 63:69--98.

\bibitem[{Head} et~al., 2002]{heameliva02}
{Head}, J.~N., {Melosh}, H.~J., and {Ivanov}, B.~A. (2002).
\newblock {Martian Meteorite Launch: High-Speed Ejecta from Small Craters}.
\newblock {\em Science}, 298:1752--1756.

\bibitem[{Housen} and {Holsapple}, 2003]{houhol03}
{Housen}, K.~R. and {Holsapple}, K.~A. (2003).
\newblock {Impact cratering on porous asteroids}.
\newblock {\em Icarus}, 163:102--119.

\bibitem[{Ivanov}, 2004]{iva04}
{Ivanov}, A.~V. (2004).
\newblock {Is the Kaidun Meteorite a Sample from Phobos?}
\newblock {\em Solar System Research}, 38:97--107.

\bibitem[{Jacobson}, 2010]{jac10}
{Jacobson}, R.~A. (2010).
\newblock {The Orbits and Masses of the Martian Satellites and the Libration of
  Phobos}.
\newblock {\em AJ}, 139:668--679.

\bibitem[{Jacobson} and {Lainey}, 2014]{jaclai14}
{Jacobson}, R.~A. and {Lainey}, V. (2014).
\newblock {Martian satellite orbits and ephemerides}.
\newblock {\em Plan. Space Sci.}, 102:35--44.

\bibitem[{Le Feuvre} and {Wieczorek}, 2011]{lefwie11}
{Le Feuvre}, M. and {Wieczorek}, M.~A. (2011).
\newblock {Nonuniform cratering of the Moon and a revised crater chronology of
  the inner Solar System}.
\newblock {\em Icarus}, 214:1--20.

\bibitem[{Michikami} et~al., 2007]{micmorhas07}
{Michikami}, T., {Moriguchi}, K., {Hasegawa}, S., and {Fujiwara}, A. (2007).
\newblock {Ejecta velocity distribution for impact cratering experiments on
  porous and low strength targets}.
\newblock {\em Plan. Space Sci.}, 55:70--88.

\bibitem[{Michikami} et~al., 2008]{micnakhir08}
{Michikami}, T., {Nakamura}, A.~M., {Hirata}, N., {Gaskell}, R.~W., {Nakamura},
  R., {Honda}, T., {Honda}, C., {Hiraoka}, K., {Saito}, J., {Demura}, H.,
  {Ishiguro}, M., and {Miyamoto}, H. (2008).
\newblock {Size-frequency statistics of boulders on global surface of asteroid
  25143 Itokawa}.
\newblock {\em Earth, Planets, and Space}, 60:13--20.

\bibitem[{Murchie} et~al., 1991]{murbrihea91}
{Murchie}, S.~L., {Britt}, D.~T., {Head}, J.~W., {Pratt}, S.~F., {Fisher},
  P.~C., {Zhukov}, B.~S., {Kuzmin}, A.~A., {Ksanfomality}, L.~V., {Zharkov},
  A.~V., {Nikitin}, G.~E., {Fanale}, F.~P., {Blaney}, D.~L., {Bell}, J.~F., and
  {Robinson}, M.~S. (1991).
\newblock {Color heterogeneity of the surface of Phobos - Relationships to
  geologic features and comparison to meteorite analogs}.
\newblock {\em J. Geophys. Res.}, 96:5925--5945.

\bibitem[Murray and Dermott, 1999]{murder99}
Murray, C. and Dermott, S. (1999).
\newblock {\em Solar System Dynamics}.
\newblock Cambridge University Press, Cambridge.

\bibitem[{Nayak} et~al., 2016]{naynimudr16}
{Nayak}, M., {Nimmo}, F., and {Udrea}, B. (2016).
\newblock {Effects of mass transfer between Martian satellites on surface
  geology}.
\newblock {\em Icarus}, 267:220--231.

\bibitem[{O'Keefe} and {Ahrens}, 1985]{okeahr85}
{O'Keefe}, J.~D. and {Ahrens}, T.~J. (1985).
\newblock {Impact and explosion crater ejecta, fragment size, and velocity}.
\newblock {\em Icarus}, 62:328--338.

\bibitem[{Pieters} et~al., 2014]{piemurtho14}
{Pieters}, C.~M., {Murchie}, S., {Thomas}, N., and {Britt}, D. (2014).
\newblock {Composition of Surface Materials on the Moons of Mars}.
\newblock {\em Plan. Space Sci.}, 102:144--151.

\bibitem[{Ramsley} and {Head}, 2013]{ramhea13b}
{Ramsley}, K.~R. and {Head}, J.~W. (2013).
\newblock {Mars impact ejecta in the regolith of Phobos: Bulk concentration and
  distribution}.
\newblock {\em Plan. Space Sci.}, 87:115--129.

\bibitem[{Ramsley} and {Head}, 2015]{ramhea15}
{Ramsley}, K.~R. and {Head}, J.~W. (2015).
\newblock {The Secondary Impact Spike of Phobos from Stickney Crater Ejecta}.
\newblock In {\em Lunar and Planetary Science Conference}, volume~46 of {\em
  Lunar and Planetary Science Conference}, page 1201.

\bibitem[{Rivkin} et~al., 2002]{rivbrotri02}
{Rivkin}, A.~S., {Brown}, R.~H., {Trilling}, D.~E., {Bell}, J.~F., and
  {Plassmann}, J.~H. (2002).
\newblock {Near-Infrared Spectrophotometry of Phobos and Deimos}.
\newblock {\em Icarus}, 156:64--75.

\bibitem[Roy, 1978]{roy78}
Roy, A.~E. (1978).
\newblock {\em Orbital Motion}.
\newblock Adam Hilger Ltd., Bristol.

\bibitem[{Ryan} and {Melosh}, 1998]{ryamel98}
{Ryan}, E.~V. and {Melosh}, H.~J. (1998).
\newblock {Impact Fragmentation: From the Laboratory to Asteroids}.
\newblock {\em Icarus}, 133:1--24.

\bibitem[{Schmedemann} et~al., 2015]{schmiciva15}
{Schmedemann}, N., {Michael}, G.~G., {Ivanov}, B.~A., {Murray}, J.~B., and
  {Neukum}, G. (2015).
\newblock {The age of Phobos and its largest crater, Stickney}.
\newblock {\em Plan. Space Sci.}, 102:152--1763.

\bibitem[Standish, 1998]{sta98}
Standish, E.~M. (1998).
\newblock Planetary and lunar ephemerides {DE405/LE405}.
\newblock Technical report, NASA Jet Propulsion Laboratory.

\bibitem[{Stuart} and {Binzel}, 2004]{stubin04}
{Stuart}, J.~S. and {Binzel}, R.~P. (2004).
\newblock {Bias-corrected population, size distribution, and impact hazard for
  the near-Earth objects}.
\newblock {\em icarus}, 170:295--311.

\bibitem[{Yoder}, 1995]{yod95}
{Yoder}, C.~F. (1995).
\newblock {Astrometric and Geodetic Properties of Earth and the Solar System}.
\newblock In {Ahrens}, T.~J., editor, {\em Global Earth Physics: A Handbook of
  Physical Constants}.

\end{thebibliography}

\end{document}